\documentclass[11pt,report]{article}

\usepackage{lineno,hyperref}
\usepackage[T1]{fontenc}
\usepackage[utf8]{inputenc}
\usepackage{authblk}
\usepackage{graphicx}
\usepackage{float}
\providecommand{\keywords}[1]{\textbf{\textit{Keywords:}} #1}
\usepackage{color}
\usepackage{soul}

\begin{document}

	\title{Effect of fluorine patterns on electronic transport in fluorinated graphene}

	\author[1,*]{Ruslan D. Yamaletdinov}
	\author[2]{Vsevolod L. Katkov}
	\author[3]{Yaroslav A. Nikiforov}
	\author[1,3]{Alexander V. Okotrub}
	\author[2]{Vladimir A. Osipov}
	
\affil[1]{Nikolaev Institute of Inorganic Chemistry, Novosibirsk, Russia}
\affil[2]{Novosibirsk State University, Novosibirsk, Russia}
\affil[3]{Bogoliubov Laboratory of Theoretical Physics, Joint Institute for Nuclear Research, Dubna, Russia}
\affil[*]{Corresponding author: Ruslan Yamaletdinov \texttt{yamaletdinov@niic.nsc.ru}}
\maketitle

	\begin{abstract}
	Within the framework of stochastic reactive molecular dynamics simulations we develop a statistical method for generating fluorinated graphene structures with desirable fluorine distribution. Electronic transport properties of fluorinated graphene in a wide range of functionalization degree and system ordering are investigated. We have found a strong correlation between irregularities in fluorine distribution and electronic  properties. In particular, proposed consideration allows us to reproduce both the experimentally observed electron-hole asymmetry in transport properties of fluorinated graphene and a recently revealed conductivity peak at 10 \%  fluoride content. 
		\end{abstract}


\section{Introduction}

Functionalization of graphene is one of the key directions in graphene research. The chemical functionalization noticeably alters the electronic structure and conductivity of graphene and introduces defect sites thereby significantly affecting the possible practical applications in optics, electronics, electrochemistry, mechanics, etc.~\cite{Feng2016}. In particular, 
 fluorine adsorption on graphene introduces $sp^3$ defects, which break the sublattice (AB) symmetry and distort the initial lattice and, as a consequence, graphene becomes  a good insulator~\cite{Osofsky2016}.
This is true at high fluorine coverage. At low and middle concentrations the situation is more delicate: localized electronic states may occur in a narrow energy window around the neutrality point. Notice that similar states emerges in hydrogenated graphene~\cite{Bang2010} where hydrogen atoms are the source of short-range disorder. The energy window grows with increasing adsorbent concentration. Experimental studies of fluorinated graphene show that the electron transport is really carried out via localized electron states in a wide range of the gate voltage~\cite{Withers2010, Withers2011, Hong2012, Tahara2013}. 

There are indications that, depending on the method of synthesis and synthesis conditions, fluorine is able to shape the different island and chain-like clusters~\cite{Vyalikh2013,Asanov2013,Pinakov2018,Boukhvalov2016}. This fact creates noticeable difficulties in theoretical description of the covalently modified graphene, since it is necessary to correctly take into account the real arrangement of functional groups.
The most popular approaches to solve this problem include random groups
distribution and periodic patterns (see, e.g., Refs.\cite{Yang2019,Makarova2017}). In fact, this is not enough to describe the real structures as evidenced by the discrepancy with the experiment. In our opinion, the main drawback of these approaches is that there is no accounting for the influence of functional groups on the activity of neighboring carbon atoms. More accurate results can be obtained by using the reactive force field (ReaxFF) based molecular dynamics (MD) simulations. The accuracy of this method was recently demonstrated by modeling the graphene oxide structure~\cite{Qiao2019}. Unfortunately this approach requires large computational power. 

In this paper, we propose a simple statistical method that allows us to generate different structures of fluorinated graphene. The distribution of fluorine atoms obtained by this low cost method is similar to the results of ReaxFF MD. 
The influence of different kinds of disorder on electronic transport in graphene has been extensively studied~\cite{Peres2010,Mucciolo2010,Sarma2011, Roche2012}. It is interesting to analyze the impact of 
chains and islands in fluorinated graphene on transport characteristics. Here we perform a series of numerical simulations to investigate the electron conductivity of two-side fluorinated graphene taking into account real arrangements of fluorine atoms at various concentrations in the range of 1-25 percent.


\section{Computational Details}
\subsection{Molecular dynamics simulations}
In this study, we use LAMMPS~\cite{plimpton1995} for MD simulations. The ReaxFF force field ~\cite{Singh2013} is used to describe the {C---F} interaction, and the charge equilibrium method QEq~\cite{Aktulga2012} is used to describe the interaction of charges in this force field.

The integration of the equations of motion in the simulations is carried out using the NVE integrator with a time step of 1 fs for 150,000 steps. The temperature is kept at 900 K using a Langevin thermostat with a dumping parameter of 10 fs. Before each simulation geometry is optimized for 2,000 time steps.

Simulations are carried out in a cell with sides $ x = 160$ \r{A}, $y=160$ \r{A}, $z=220$ \r{A}. In the center of the $z$ axis on the $xy$ plane a square graphene sheet with sides $L = 100$ \r{A} is placed, and the corners of the sheet are fixed. Fluorine  molecules ($\textrm{F}_2$) are randomly located in a cell. The  number of $\textrm{F}_2$ varies from 2500 to 4500 molecules per cell with increment of 500 molecules per cell. For each of these cases, 100 simulations with different randomized initial arrangement of fluorine molecules are performed.

\subsection{Statistical method}

In order to take into account the more realistic arrangement of fluorine atoms we suggest a rather simple method based on statistical consideration. The essence of the method is as follows:  at each iteration,
unfluorinated carbon atom is randomly selected. The fluorination probability of this atom $p_i$ is determined taking into account the presence of bonded fluorine on a neighboring atoms. We use the Boltzmann-like  distribution in the form
\begin{equation}
\label{eq:prob}
p_i=Z^{-1}\cdot A_i\cdot \exp{(-\beta  E_i)},
\end{equation} 
where $E_i$ is a bonding energy in the current environment, $A_i$ is an efficient availability of this carbon atom, $Z=\sum_{j}A_j\cdot \exp{(-\beta  E_j)}$ is a normalization factor (summation is carried out over all possible states), $\beta$ is considered as a free model parameter which governs ordering of fluorine atoms. $E_i$ is parameterized by five values that correspond to five possible types of interaction: bond energy, \textit{cis-} and \textit{trans-} steric interaction with closest substituents (if present), energy of $\pi$ bound and lattice disturbance energy correction. The structure averaged C---F interaction energy varies in the range of $65-80$ kcal/mol and depends on fluorination degree and ordering parameter $\beta$. $A_i$ values was optimized for best fit of our ReaxFF MD results.  A detailed description of the method, parameters and optimization processes is presented in Supporting Information.

\subsection{Transport calculation}

Accounting for the effects of structural ordering on transport properties requires a huge simulation cell and averaging over a large number of structures. As the most appropriate method we have chosen the recursive nonequilibrium Green's functions method (RGF)~\cite{Lewenkopf2013}  based on the  tight-binding Hamiltonian~\cite{Yuan2015}. This method allows one to iteratively solve the transport problem for large structures and requires only specifying the correct Hamiltonian of a system~\cite{Katkov2017} (detailed  description of the modern methodologies for quantum transport can also be found in a recent review~\cite{Fan2018}). 

\begin{figure*}[!h]
	\begin{center}
	\includegraphics[width=\textwidth]{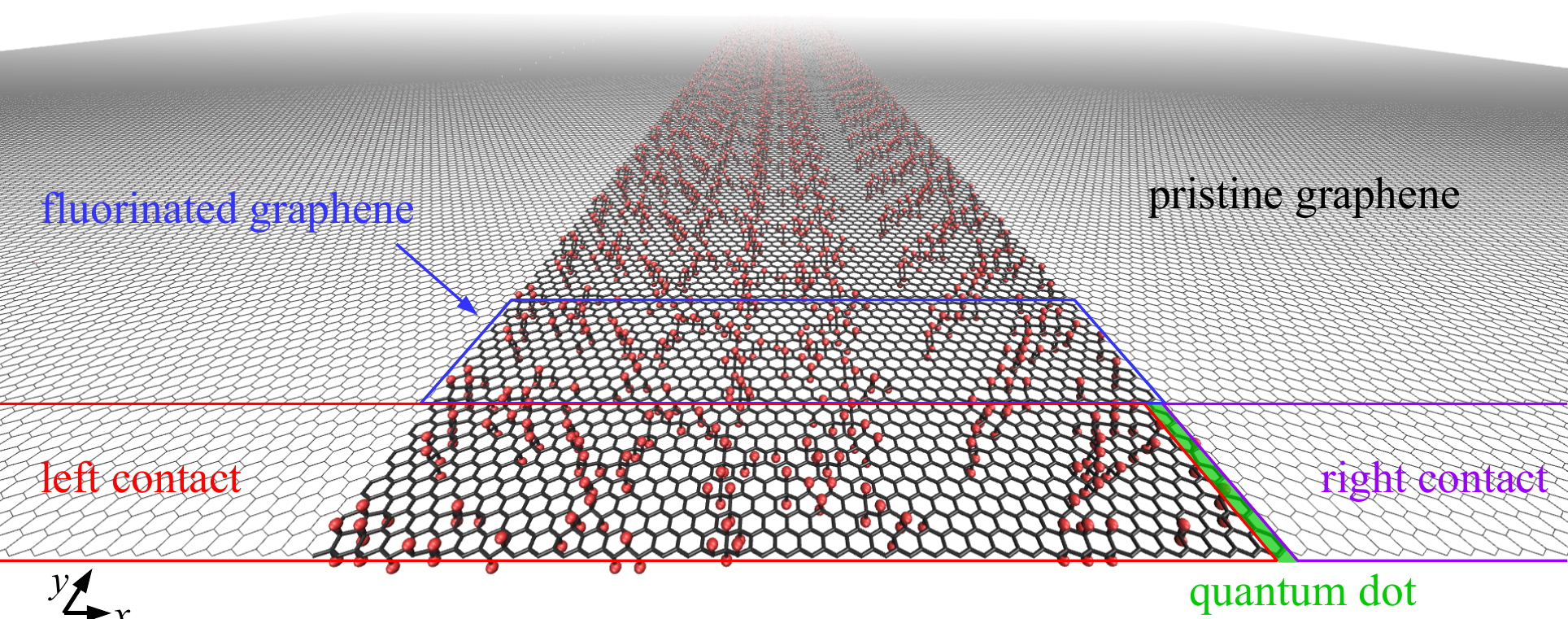}
	\caption{Model system for transport calculations. Fluorinated graphene (blue rectangle) is situated between two semi-infinity graphene sheets with periodic boundary conditions along the $y$-axis. On the bottom of the figure the final iteration is demonstrated.}
	\label{fig:ssetup}
	\end{center}
	
\end{figure*}

As a model system we consider the contact/quantum dot/contact geometry with periodic boundary conditions along the $y$-axis and contacts located along armchair edges. At every iteration the semi-infinite pristine graphene sheet is used as a right contact (see magenta rectangle in Fig.~\ref{fig:ssetup}). The left contact and quantum dot are used to be iterable. At the first iteration the semi-infinity pristine graphene sheet is regarded as a left contact and the first armchair line of fluorinated part as a quantum dot. In the $N$-th step the left semi-infinity graphene sheet together with $N-1$ left armchair lines of fluorinated graphene are used as a left contact, and $N$-th armchair line is used as a quantum dot. In Fig.~\ref{fig:ssetup} both the contacts and quantum dot after the last iteration are highlighted.

In order to reveal the transport properties depending on the amount of fluorine in the structure ($x$ in $\mathrm{CF}_x$) and the parameter $\beta$ we perform a series of calculations. Our studies show that $\mathrm{CF}_x$ piece of $21.32$~\r{A} wide (along armchair edge) and $36.92$~\r{A} long (along zigzag edge) is optimal for numerical calculations. For each value of $x$ and $\beta$,  $10^2$ calculations were carried out, followed by averaging.   
 
\section{Results and discussion}

\begin{figure*}[h]
	\begin{center}
		
		\begin{minipage}{0.32\textwidth}
			\includegraphics[width=1\linewidth]{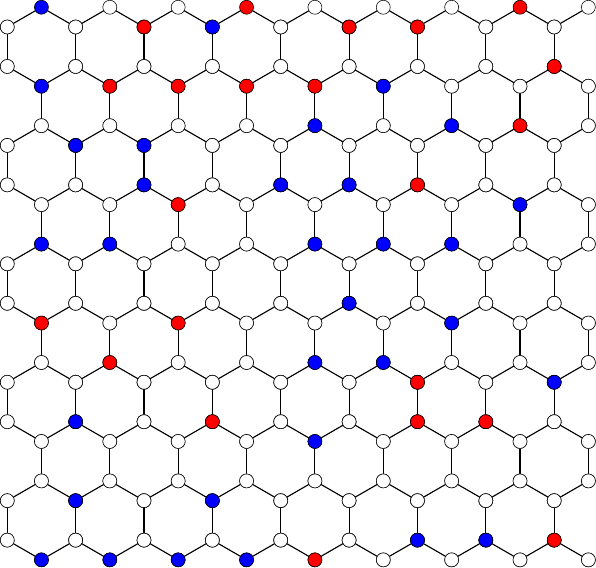}\\(a)
		\end{minipage}
		\hfill
		\begin{minipage}{0.32\textwidth}
			\includegraphics[width=1\linewidth]{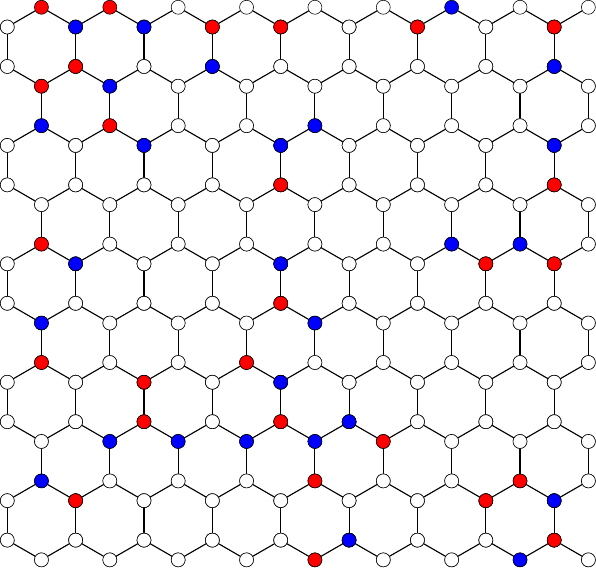}\\(b)
		\end{minipage}
		\hfill
		\begin{minipage}{0.32\textwidth}
			\includegraphics[width=1\linewidth]{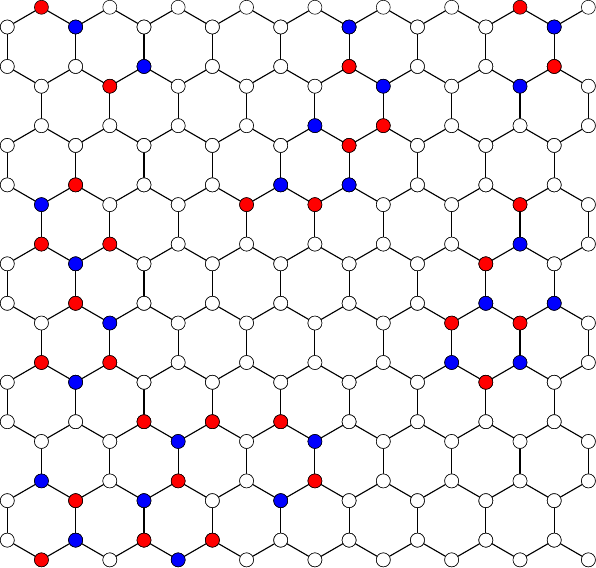}\\(c)
		\end{minipage}
		\caption{Generated structures of $CF_{0.3}$ at different values of $\beta$: (a) $\beta=0$ mol/kcal  (random distribution), (b) $\beta=0.1$ mol/kcal, and (c) $\beta=0.2$ mol/kcal. White circles denote carbons atoms, red and blue circles denote up and down oriented F atoms.}
		\label{fig:structs}
	\end{center}
\end{figure*}

Examples of structures generated by our method are presented in Fig.~\ref{fig:structs}. It is known that fluorine patterns on graphene tend to form branched chains and nanoislands~\cite{Vyalikh2013,Asanov2013,Pinakov2018,Boukhvalov2016}. Within the framework of our method, the size of agglomerated areas is directly related to $\beta$: higher $\beta$ values lead to more structure ordering. This means that longer chains and larger islands appear at the same fluorine concentration. It is clearly visible in Fig.~\ref{fig:structs}: when $\beta=0$ mol/kcal  the random distribution is observed (Fig.~\ref{fig:structs} (a)), while for $\beta=0.1$ mol/kcal one can see the formation of branched chain (Fig.~\ref{fig:structs} (b)) with further agglomeration of chains into islands at $\beta=0.2$ mol/kcal (Fig.~\ref{fig:structs} (c)). From an experimental point of view such ordering growth may correspond to a decrease in reaction temperature~\cite{Wang2012}. 

There are various methods to synthesize fluorinated graphene like exfoliation of graphite fluoride~\cite{Zboril2010}, hydrothermal reduction of graphene oxide with HF~\cite{Wang2012}, and direct graphene fluorination utilizing $\mathrm{XeF}_2$~\cite{Nair2010},   $\mathrm{Br}_2/\mathrm{BrF}_3$~\cite{Pinakov2006}, and plasma ($\mathrm{CF}_4$, $\mathrm{SF}_6$ and $\mathrm{Ar/F}_2$)~\cite{Feng2016}. It is clear that in case of exfoliation or reduction the final fluorinated graphene pattern is determined by precursor structure, so our model can be used only for the  direct graphene fluorination process. In this case, the parameter $\beta$ can be correlated with synthesis conditions such as temperature, pressure, free radicals concentrations, etc.: the stricter the synthesis conditions, the lower $\beta$ should be taken. In order to verify the statistical method we compared our  results with ReaxFF MD simulations and found that the best fit is reached at $\beta=0.12$ mol/kcal. Thus, we can state that our model can reproduce at least qualitatively the experimental observations. 

\begin{figure}[!h]
	\begin{center}
		
		\begin{minipage}{0.49\textwidth}
			\includegraphics[width=1\linewidth]{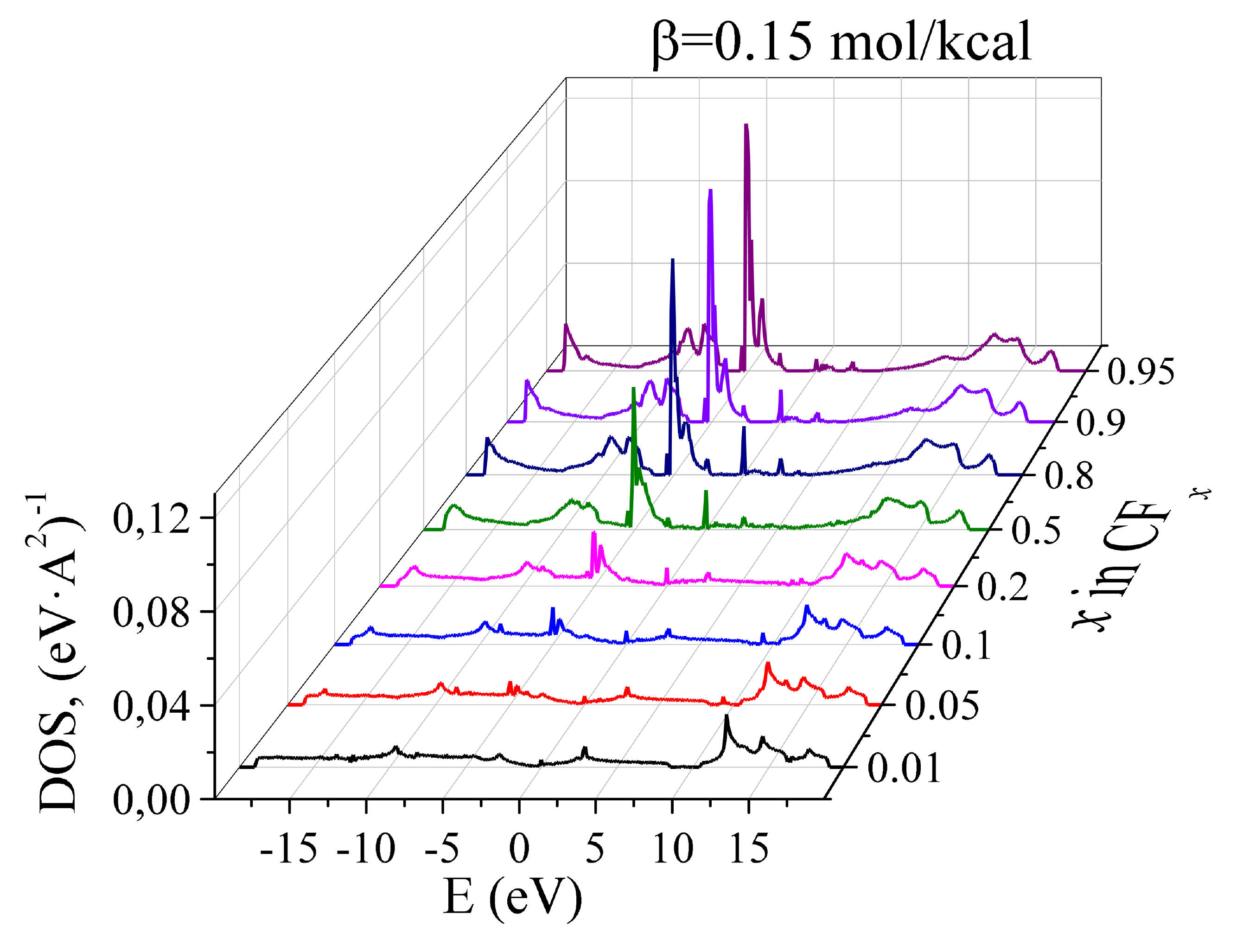}\\(a)
		\end{minipage}
		\hfill
		\begin{minipage}{0.49\textwidth}
			\includegraphics[width=1\linewidth]{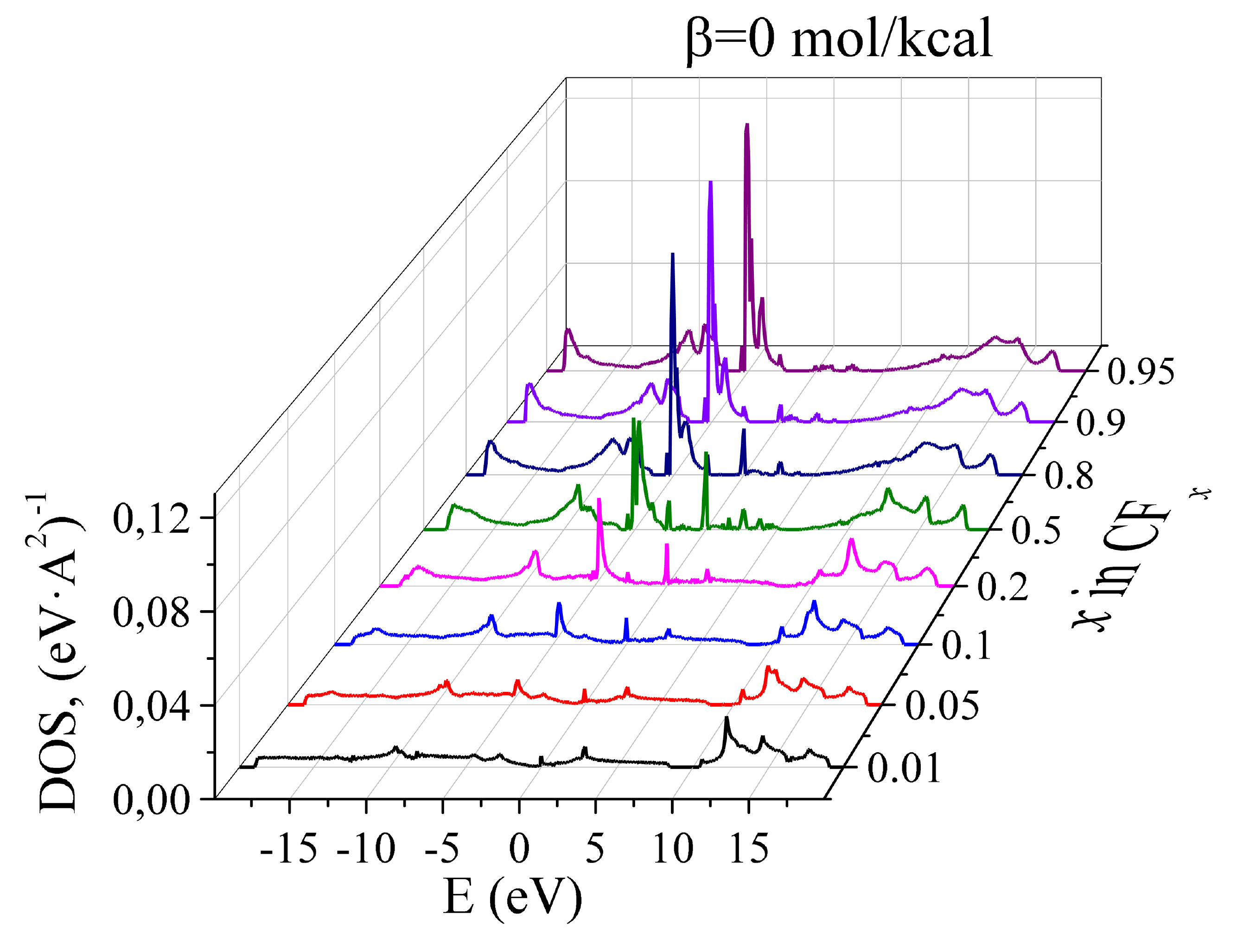}\\(b)
		\end{minipage}
		
		\caption{DOS of fluorinated graphene $\textrm{CF}_x$ at different $x$ for two values of $\beta$:  $\beta=0.15$~mol/kcal (a) and $\beta=0$ (b).}
		\label{fig:dos}
	\end{center}
\end{figure}

Calculations of transport properties were performed for structures generated by our statistical method.
As a first step we calculated the density of electronic states (DOS) at different  $x$  and $\beta$ (see Fig.~\ref{fig:dos}). Notice that a sharp peak near the Fermi energy ($E\approx0$~eV) may be caused by midgap states of unpaired fluorine atoms that violet locally AB symmetry~\cite{Yuan2015}. With a greater ordering of the system ($\beta\neq 0$), the peak value decreases due to a decrease in the number of unpaired fluorine atoms. One can also see a slight asymmetry of DOS in the region of $\pm2$ eV associated  with carbon $p_{xy}$ orbitals. 

In our case, the electron transport in localization regime is favored for fluorinated graphene segment with a length of more than $40$~\r{A} and in the energy range of $\pm1.2$~eV. Performing statistical averaging we calculated the localization length $\xi$ as a function of energy, fluorine fraction $x$ and $\beta$. More precisely, the localization length was calculated by the exponential fit of the $\textrm{CF}_x$ conductivity: $\sigma_{2D}(l)\propto \exp (- l/\xi)$, where $\sigma_{2D} = 2 e^2 l/ h \times \langle T\rangle$ is 2D conductivity as a function of length $l$ and $\langle T\rangle$ is the transmission averaged over all disordered configurations ($e$ is the elementary charge, and $h$ is the Planck constant). We also verified known localization conditions~\cite{Avriller2007, Cresti2009} in the form $\Delta T/\langle T\rangle>1$ and $\Delta  \ln T/\langle\ln T\rangle<1$, where $\Delta$  means the standard deviation. 

It was earlier found~\cite{Withers2011,Tahara2013} that temperature-dependent resistance of fluorinated graphene is well fitted by 2D variable-range hopping model (2D-VRH) with $\rho=\rho_0\cdot\exp \left[(T_0/T)^{1/3}\right]$, where  $k_B T_0=13.6/\xi^2 \rho(\varepsilon_F)$, $\rho(\varepsilon_F)$ is the density of localized states at the Fermi level, $\rho_0$ is a prefactor, $k_B$ is the Boltzmann constant, $T$ is a temperature, and $\xi$ is the localization length~\cite{Shklovsky1984}.
To estimate the value of $T_0$ we used previously fitted $\xi$ and calculated density of states at given energy (Fig.~\ref{fig:dos}) assuming that most of them are localized in the studied energy range $(-1.2;~1.2$)~eV.

\begin{figure}[!h]
		\begin{minipage}{0.49\textwidth}
			\includegraphics[width=1\linewidth]{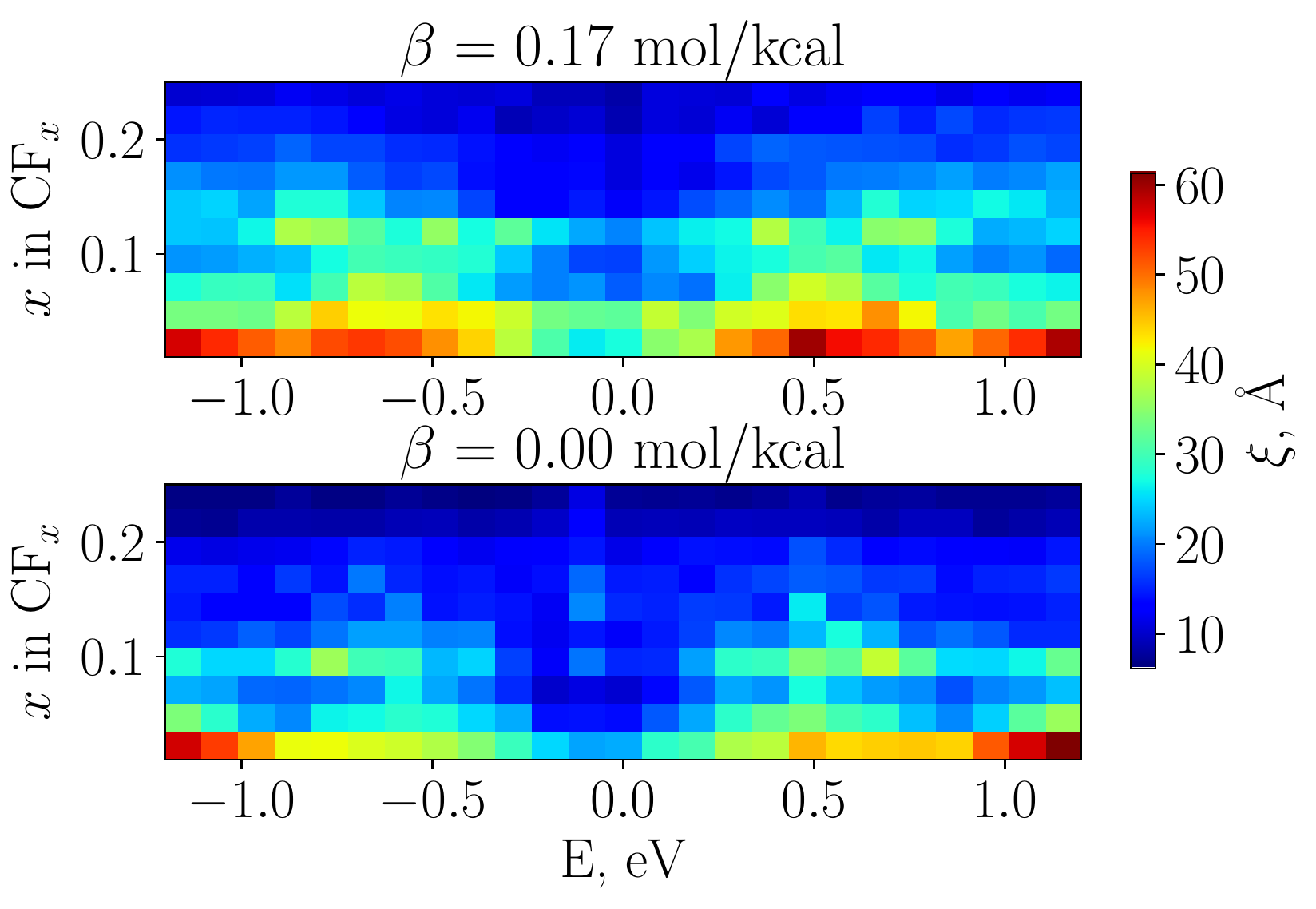}\\(a)
		\end{minipage}
		\hfill
		\begin{minipage}{0.49\textwidth}
			\includegraphics[width=1\linewidth]{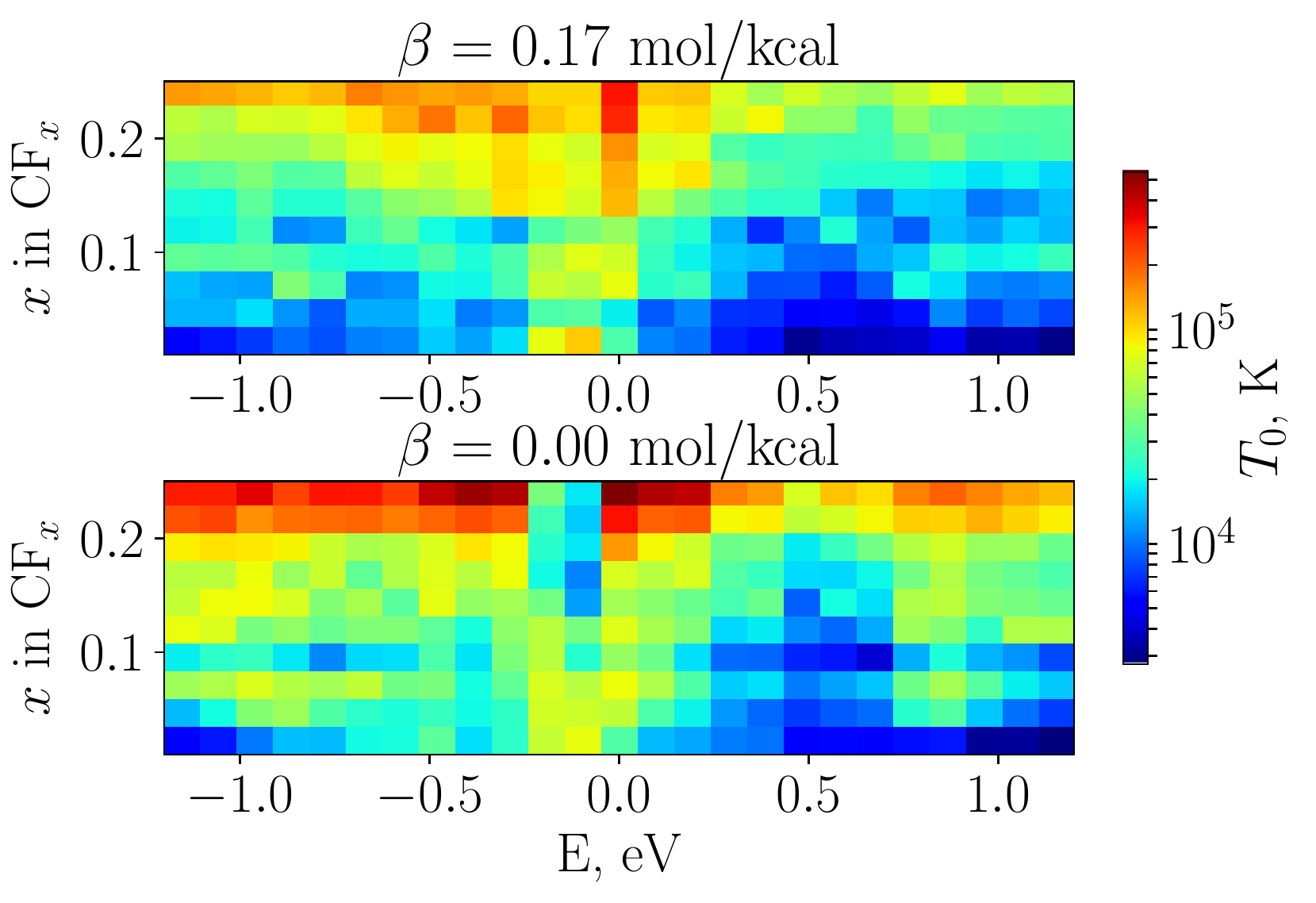}\\(b)
		\end{minipage}
		\caption{Localization length $\xi$ (a) and 2D-VRH temperature $T_0$ (b) as a function of fluorination $x$, energy and the parameter $\beta$.}
		\label{fig:LT}
\end{figure}

Localization length as a function of the energy and the fluorine concentration is shown in  Fig.~\ref{fig:LT} (a). Notice that the $\xi$ values have the same order of magnitude as for the disordered hydrogen atoms~\cite{Bang2010} at the same concentration.
The calculations were performed for $\beta = 0$ and 0.17~mol/kcal which corresponds to random and ordered distribution of fluorine, respectively. The concentration axis starts from 1 \% ($x$=0.01). At this and higher concentration the above localization conditions are well satisfied. 

As seen in Fig.~\ref{fig:LT}, the localization length significantly depends on the type of adatom distribution. In both cases, $\xi$ is located quite symmetrically about the vertical axis. As could be expected, the localization length decreases with increasing concentration. Surprisingly, our calculations revealed a maximum of $\xi$ at a fluorine concentration of about 10\% ($x\approx 0.1$). This maximum is more pronounced in case of $\beta = 0.17$~mol/kcal in the region near the Fermi level. It should be noted that a similar maximum in conductivity has been recently found experimentally in thin films of few-layer graphene with fluorine content at the same concentration~\cite{Kolesnik2018}. For $\beta = 0$~mol/kcal this maximum moves to higher energies $|\varepsilon|>0.2$ eV. Since the transport measurements in~\cite{Kolesnik2018} were performed at zero gate voltage, we can conclude that the appearance of conductivity maximum  at 10\%  is a direct consequence of a more ordered distribution of fluorine atoms.

The values of the hopping parameter $T_0$ as a function of the energy and concentration are shown in Fig.~\ref{fig:LT} (b). In both cases, the calculated order of magnitude and revealed asymmetric behaviour are in good agreement with existing experimental results ($T_0\sim10^2-10^5$ K ~\cite{Withers2011,Tahara2013}). Nonetheless, the case of ordered distribution gives more adequate coincidence. One can see a peak at the Fermi level   and a long tail in the hole region with the presence of additional small  sub-peaks (one or two) depending on the degree of fluorination. Taking into account that the localization length is quite symmetrical in  Fig.~\ref{fig:LT} (a) one can conclude that the origin of asymmetrical behavior of $T_0$ is the asymmetry in DOS (see Fig.~\ref{fig:dos}). The calculated 2D-VRH conductivity at 300~K shows a pretty similar to measured trend with a smooth decrease with increasing fluorination degree up to 2 orders of magnitude and a local conductivity maximum for $\textrm{CF}_{0.1}$ structure.

\begin{figure}[!h]
		\begin{minipage}[b]{0.4\textwidth}

			\includegraphics[width=1\linewidth]{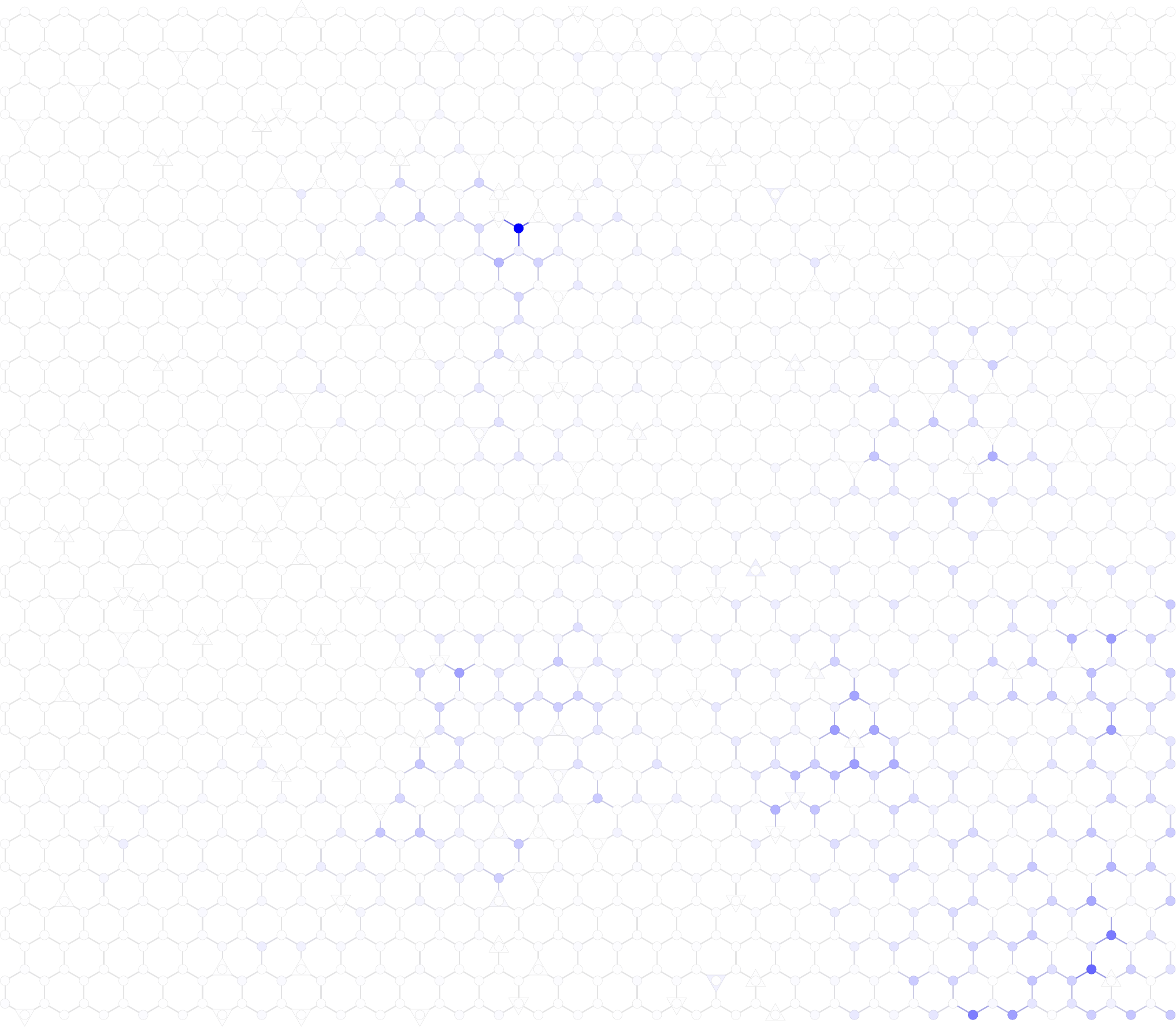}\\(a)
		\end{minipage}
		\hfill
		\begin{minipage}[b]{0.55\textwidth}

			\includegraphics[width=1\linewidth]{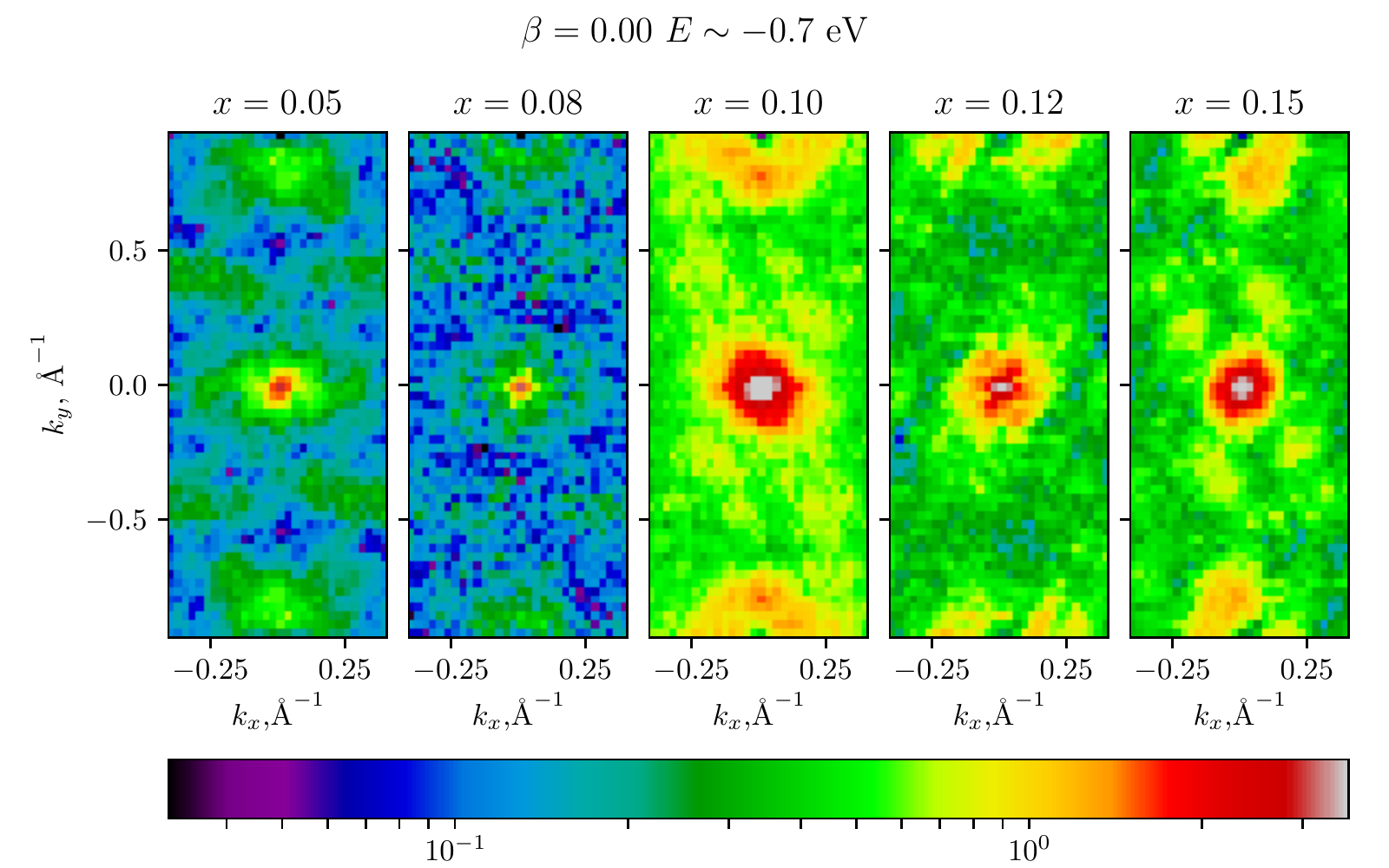}\\(b)
		\end{minipage}
		\caption{Electron density of a localized state at the energy $E\sim -0.7$~eV for $\textrm{CF}_{0.08}$ and $\beta=0$ (a). The fourier transform of the electron density at the energy  $E\sim -0.7$~eV and $\beta=0$ for different $x$ in $\textrm{CF}_x$ (b).}
		\label{fig:edens}
\end{figure}

It is known that the rising disorder leads to the formation  of electron domains in which charge carriers are localized. Based on our results, we can assume that the size and position of electronic domains meet the resonance tunneling conditions at the fluorine concentration of about 10 percent. In this case, 
electrons can spread over several domains thus increasing the effective size of the delocalization which in turn will lead to drastic changes in transport properties. To confirm this assumption we calculated the charge density at given values of energy. The charge density near E$\sim -0.7$~eV for $\textrm{CF}_{0.08}$ at $\beta=0$ is presented in Fig.~\ref{fig:edens}(a). As is seen, the specific spotted patterns are quite typical for fluorinated structures in an energy  range of $\pm2$~eV. For resonance tunneling we would expect the formation of quasiperiodic electron structures. Periodicity can be detected by Fourier analysis. We made it and found a series of apparent peaks with a wavelength of $1.4$~\r{A} to $12$~\r{A} (see Fig.~\ref{fig:edens}(b) $k_y$ aligned along armchair edge, $k_x$ - along zig-zag edge). It is easy to see the obvious difference between $\textrm{CF}_{0.08}$ and  $\textrm{CF}_{0.1}$, especially in the region $\vec{k}\approx0$. Unfortunately, it is impossible to resolve the central region with $\vec{k}\approx0$ more accurately due to a large number of relevant electronic domain structures.

\section{Conclusion}
We offer new statistical method to simulate the structure of fluorinated graphene, which can be tuned for structure generation with desirable ordering and accurately meets ReaxFF MD simulations results.
We calculated the density of states, localization length and 2D-VRH characteristic temperature $T_0$ for highly fluorinated ghaphene with two types of disorder distribution. We found that electrons for considered fluorine concentrations (1-25\%) exhibit localized behavior in a wide range of energies close to the neutrality point.
The transport simulations allow us to reproduce (i) the recently found conductivity peak at 10 \% fluoride content ~\cite{Kolesnik2018} associated with the emergency of quasi-periodical structures in electron density and (ii) experimentally observed electron-hole asymmetry in fluorinated graphene~\cite{Tahara2013} which can be connected with specific patterns in the density of states. 
We have shown that transport properties of fluorinated graphene are strongly correlated with system ordering, which confirm that account of real fluorine distribution  is important for modelling the electron transport phenomena in such systems.

\section*{Acknowledgment}
The reported study was funded by RFBR according to the research project N. 19-32-50013.

\renewcommand{\thefigure}{S.\arabic{figure}}
\setcounter{figure}{0}
\renewcommand{\thetable}{S.\arabic{table}}
\setcounter{table}{0}
\renewcommand{\theequation}{S.\arabic{equation}}
\setcounter{equation}{0}
\section{Supporting Information}
\subsection{Statistical model}

As it mentioned in the main text,  our model works as follows:  at each iteration,
unfluorinated carbon atom is randomly selected. The fluorination probability of this atom $p_i$ is determined taking into account the presence of bonded fluorine on a neighboring atoms. We use the Boltzmann-like  distribution in the form
\begin{equation}
\label{eq:probb}
p_i=Z^{-1}\cdot A_i\cdot \exp{(-\beta  E_i)},
\end{equation} 
where $E_i$ is a bonding energy in current environment, $A_i$ is an efficient availability of this carbon atom, $Z=\sum_{j}A_j\cdot \exp{(-\beta  E_j)}$ is a normalization factor (summation is carried out over all possible states), $\beta$ is considered as a free model parameter which governs ordering of fluorine atoms. $E_i$ is parameterized by five values that correspond to five possible types of interaction: bond energy, \textit{cis-} and \textit{trans-} steric interaction with closest substituents (if present), energy of $\pi$ bound and lattice disturbance energy correction (see table~\ref{tab:E_decomp}). 
\begin{table}[!h]
	\begin{center}
		\caption{Terms of $E_i$ decomposition (in kcal/mol).}
		\label{tab:E_decomp}
		\begin{tabular}{|c|c|c|c|c|}
			\hline
			$E_{bond}$&$E_{cis}$&$E_{trans}$& $E_\pi$& $E_{single}$ \\
			\hline	 		
			-89.3&18.0 & 5.1&4.5&18.9\\
			
			\hline
		\end{tabular}
	\end{center}
\end{table}
In order to parameterize $E_i$ we calculate a C-F bond energy for different surroundings. According to our MD calculations there are 7 stable conformations: single F atom without any fluorinated neighbors (\textit{0}), F atom with one \textit{cis-}orientated neighbor (\textit{c}), F atom with one \textit{trans-}orientated neighbor (\textit{t}), F atom with one \textit{cis-} and one \textit{trans-}orientated neighbors (\textit{ct}),  F atom with two \textit{trans-}orientated neighbors (\textit{tt}), F atom with one \textit{cis-} and two \textit{trans-}orientated neighbors (\textit{ctt}) and F atom with three \textit{trans-}orientated neighbors (\textit{ttt}). Energy $E_i$ is calculated in depends of neighbors orientation.
 
\begin{table*}[!h]
	\begin{center}
		\caption{Optimized $A_i$ for different orientations.}
		\label{tab:A_i}
		\begin{tabular}{|c|c|c|c|c|c|c|c|}
			\hline
			neighbors orientations&\textit{0}&\textit{c}&\textit{t}&\textit{ct}&\textit{tt}&\textit{ctt}&\textit{ttt} \\
			\hline	 		
			$A_i$&6.47&0&1.43&2.15&0.40&1.24&1.49   \\
			
			\hline
		\end{tabular}
	\end{center}
\end{table*}

To optimize $A_i$ values we fit 500 ReaxFF MD simulations results. The deviation of the results was calculated based on the statistics of different orientations (\textit{0},\textit{c},\textit{t},\textit{ct},\textit{tt},\textit{ctt},\textit{ttt},) for each F atom. Optimized $A_i$ values are presented in table~\ref{tab:A_i}.

\bibliographystyle{ieeetr}

\bibliography{biblio}

\end{document}